# Unequal Rewards and the Direction of Scientific Work: Evidence from U.S. Public Universities

Likun Cao, Department of Sociology, Purdue University. E-mail: lkcao@purdue.edu

**Abstract**

Research on pay dispersion has long asked whether unequal rewards increase or depress performance. Tournament theory expects stronger effort, while social comparison accounts expect withdrawal. Both perspectives, however, treat performance as a single scale on which workers can be ranked. Scientific work often resists such ordering. Some contributions extend recognized lines of inquiry and can be evaluated against prevailing standards; others revise the frames through which value is judged and become legible only later. We argue that intellectual tournaments more readily reward the first kind of contribution than the second. Pay dispersion may therefore be associated not only with how much scholars produce but also with what kinds of intellectual exploration are encouraged. We test this argument with administrative payroll records for 105,050 tenure-line faculty in 4,839 departments at 93 U.S. public universities from 2004 to 2026, linked to publication histories in OpenAlex. For each department, we estimate a separate pay equation using rank, contract basis, and seniority, then measure the Gini of the salary variation this local ladder cannot explain. Scholars in departments with greater unexplained pay dispersion publish more papers and receive more citations, but their work is less disruptive, less novel, and more conventional. Additional analyses show that compensation premiums above the ladder go disproportionately to scholars with broad collaborations and topical breadth. Although the design cannot fully separate treatment from selection, the evidence shows that inequality is not only a distributional outcome but also an evaluative structure, shaping which forms of knowledge are recognized, rewarded, and reproduced within organizations and communities.

**Acknowledgments.** The author thanks Jiao (Lorena) Ning for assistance with data collection and cleaning. The author takes full responsibility for the content of this article.

## 1. Introduction

Inequality inside organizations is produced by the ordinary practices through which employers attach pay to people and positions (Avent-Holt and Tomaskovic-Devey 2019; Tomaskovic-Devey and Avent-Holt 2019). Consider two academic departments, both ordinary. In one, two colleagues drawn at random differ in salary by roughly $16,000 once the department's ladder of rank, contract basis, seniority, and year has been accounted for. In the other, the same two colleagues differ by roughly $53,200. Both departments have an official ladder, and both can explain why a full professor earns more than an assistant professor. Neither can explain the difference between two numbers on the ladder's own terms. That residual dispersion is how unequal a reward structure remains after the hierarchy everyone can see has been taken into account.

Whether such dispersion helps or harms the organization is a long-running question that attracted rich debate. In tournament theory, unequal rewards are prize spreads that motivate effort (Lazear and Rosen 1981). In the social comparison and equity traditions of sociology and social psychology, unequal rewards are relational signals that invite judgments about worth, legitimacy, and fairness. When those judgments are met with skepticism, inequality can weaken cooperation and motivation (Adams 1965; Pfeffer and Langton 1993). Four decades of research have asked which reading is right, and the literature's conclusion is that both may be, in different settings (Shaw, Gupta, and Delery 2002; Shaw 2014). This paper asks a different question: what happens when the output being rewarded is knowledge, whose value cannot be directly evaluated and is not equally legible across forms of work? Some knowledge is produced within an existing evaluative frame and can be ranked by familiar signals: productivity, community recognition, and conformity to standards in force. Other knowledge revises the frame itself, and its value depends on standards that have not yet stabilized (Kuhn 1962). A tournament, by definition, can reward the first kind more readily than the second. Pay dispersion may therefore encourage knowledge that fits existing evaluative standards while discouraging work that revises them.

We test this hypothesis with a unique dataset collected and assembled with support from a novel AI-agent pipeline. Using Claude Code for scraping support, we collected administrative payroll records for 93 U.S. public universities from public disclosures, covering fiscal years 2004 to 2026, and linked them at the individual level to publication histories in OpenAlex. The analytic sample includes 124,056 department-person records, covering 105,050 tenure-track faculty members in 4,839 departments at U.S. public universities. Pay dispersion is measured with a residualized departmental Gini. For each department, we estimate its own pay equation to remove the variation accounted for by its formal ladder. The Gini of the remaining pay captures the department's unexplained dispersion. We then relate this measure to the research its members produce across five dimensions: productivity, citations, novelty, conventionality, and disruption (Uzzi et al. 2013; Funk and Owen-Smith 2017; Wu, Wang, and Evans 2019).

Three findings emerge from the analysis. First, departments with more unexplained pay dispersion produce more publications and accumulate more citations, as tournament theory expects. However, they also produce research that is less disruptive, less novel, and more conventionally connected to prior literature. The associations survive the removal of big team scholars from every department, and they do not depend on residualizing rank: raw pay inequality and the inequality the ladder itself creates carry the same signs. Second, the individual pay premium above a department's ladder rewards team size and topical breadth, not disruption or novelty. This runs against the common view that specialization brings greater success: at least financially, the premium goes to more generalist and more collaborative scholars. Third, two tests probe where the pattern arises, although neither can separate a treatment effect of dispersion from the selection of scholars into and out of departments. More unequal departments show no statistically detectable tendency to appoint assistant professors whose prior work already differs on any of the five dimensions, so the pattern is not visibly a recruitment pattern. But when a department's top-paid senior member retires, reducing pay dispersion, the colleagues who remain become more disruptive.

The paper makes contributions to both research on organizational inequality and scientific careers. Theoretically, it shows that the two established accounts of pay dispersion are each partly right, but about different components of knowledge output. By bringing STS theories into the study of organizational inequality, we show why economic theories of incentives cannot fully explain all dimensions of creative work. Competition may increase visible output and reward work that is easier to recognize, but it is also associated with less of the more uncertain, disruptive, or paradigm-challenging forms of inquiry, a cost to innovation. Methodologically, this paper introduces a measure of within-organization inequality defined on the organization's own terms: the dispersion that remains after each unit's own reward ladder is estimated. It also demonstrates the potential of AI agents for large-scale data collection, an approach that could be extended to many other kinds of institutional and administrative records.

The design identifies a conditional association, not a causal effect: departmental pay dispersion is highly stable over time and is not assigned. This evidence is therefore not intended to establish causality, and we state these limits throughout.

The remainder of the paper proceeds as follows. Section 2 reviews the two accounts of pay dispersion, identifies the assumption they share, explains why knowledge work unsettles it, and states two hypotheses based on the arguments. Section 3 describes the data, the residualized Gini measure, other variables, and the estimation. Section 4 reports the main estimates, a comparison of raw, ladder-explained and residualized inequality, the individual premium, robustness checks, and the potential alternative explanations. Sections 5 and 6 extend our discussion and conclude.

## 2. Background and hypotheses

### 2.1. Pay dispersion as incentive and as comparison

The literature on pay dispersion begins from an ambiguity: the same distribution can be an incentive schedule or a comparison structure. In the economic reading, pay differences are prizes. Rank-order tournament theory shows that compensation need not price absolute output if workers can be ordered relative to one another, and that a wider spread between prizes gives contestants a stronger reason to exert effort (Lazear and Rosen 1981). Laboratory and field evidence is broadly consistent with the model (Bull, Schotter, and Weigelt 1987; Ehrenberg and Bognanno 1990), and promotion-based reward systems are among the most common incentive structures inside firms (Baker, Jensen, and Murphy 1988; Connelly et al. 2014).

The sociological and social-psychological tradition assigns the same distribution a different meaning. Social comparison theory treats the evaluation of one's own performance as relational (Festinger 1954); equity theory specifies the comparison as a ratio of outcomes to inputs against a referent (Adams 1965); and organizational justice research adds that people judge the procedure and legitimacy of an allocation as much as the amount received (Deutsch 1985; Lind and Tyler 1988; Colquitt et al. 2001). Dispersion experienced as unfair can elicit withdrawal, conflict, or reduced cooperation (Akerlof and Yellen 1990; Fehr and Schmidt 1999). Learning that peers earn more lowers satisfaction (Card et al. 2012), pay inequality within work groups lowers output (Breza, Kaur, and Shamdasani 2018), and large employers compress wages to blunt the comparisons their size invites (Cobb and Stevens 2017).

These two views are often tested in the same settings. Studies of executive compensation and academic departments show that pay dispersion can be linked to lower satisfaction, weaker collaboration, and lower research productivity (O'Reilly, Main, and Crystal 1988; Pfeffer and Langton 1993). Later work shows why the evidence is mixed: incentive effects and comparison effects can operate at the same time, and the consequences of pay dispersion depend partly on whether inequality occurs between ranks or within them (Shaw et al. 2002; Shaw 2014; Kacperczyk and Balachandran 2018).

The seemingly opposed claims, however, share an assumption: both require a basis of comparison. A tournament needs a signal that can rank contestants when prizes are allocated, that rises with effort, and that is not easily distorted by sabotage (Lazear and Rosen 1981; Prendergast 1999; Chen 2003). A social comparison account also assumes that members' contributions can be compared. Categories, measures, and evaluation practices make different people and outputs commensurable (Espeland and Stevens 1998). Once measures are tied to rewards, they shape attention and behavior rather than simply record them (Espeland and Sauder 2007; Sauder and Espeland 2009; Dahler-Larsen 2014; de Rijcke et al. 2016). Even when organizations claim to pay for performance, the translation of ratings into pay can introduce bias, though transparency and accountability can reduce it (Castilla 2008, 2015).

What has received less attention is the harder case: rewards for work that cannot easily be placed on a common scale.

### 2.2. Knowledge work and the limits of comparison

The assumption of comparability does not travel easily to knowledge work. Although rankings permeate professional organizations (Espeland and Sauder 2007; Sauder and Espeland 2009), knowledge output contains a stochastic component that effort does not govern. A great work is often recognized many years after it is created (Ke et al. 2015; Wang, Veugelers, and Stephan 2017), and the timing of a scientist's highest-impact contribution is approximately random within their own sequence of work (Sinatra et al. 2016).

Underlying this is a basic problem of observability in knowledge work. What colleagues or knowledge-producing organizations can observe is usually a record of output volume, peer reception, and the judgments of established members of the field: in science, publications, citations, and grant portfolios. These signals measure performance against the standards currently in force. A citation, for example, is a researcher's judgment, made within the prevailing frame, that a contribution matters to a problem the field already recognizes. Indicator systems built from such judgments therefore tend to score conformity to an existing research front.

For the same reason, highly novel work is often harder to evaluate when it first appears. Frame-revising contributions change the standards by which work is judged, and at the time of evaluation, those new standards have often not yet been consolidated. Evidence from the economics of innovation and the psychology of creativity points in the same direction: exploration requires tolerance for early failure and long time horizons, while contingent extrinsic rewards can increase output on well-specified tasks but depress performance on tasks requiring heuristic search (Amabile 1996; Frey and Jegen 2001; Lerner and Wulf 2007; Azoulay, Graff Zivin, and Manso 2011; Manso 2011; Ederer and Manso 2013).

This does not mean that the logic of comparability, which is underlying tournament and social comparison accounts, fails in knowledge work. It means something more specific and testable: tournaments and comparisons operate most strongly on the performance dimensions people can directly observe. As the multitask agency problem implies, when agents allocate effort across multiple tasks and principals observe only some of them, stronger incentives on the observed task redirect effort away from the unobserved one (Holmström and Milgrom 1991). The prediction, then, is not about the level of performance alone, but about its composition.

### 2.3. Frame-conforming and frame-revising knowledge

Kuhn (1962) provides a simple and sharp theoretical framework for distinguishing between frame-conforming and frame-revising knowledge. Normal science works within an existing paradigm; revolutionary science transforms the standards by which contributions are judged. Frame-conforming work is legible at the moment; frame-revising work becomes legible only as the standards change, after the rankings that might have rewarded it have been made. Kuhn (1977) places the same tension inside the scientist's own portfolio, and Foster, Rzhetsky, and Evans (2015) recover it in the citation record: scientists mostly choose conservative strategies, innovative strategies earn higher rewards when they succeed, and the expected return nonetheless favors conservatism.

The science of science now makes the two modes measurable at population scale (Fortunato et al. 2018; Wang and Barabási 2021). Funk and Owen-Smith (2017) propose a disruption index that distinguishes work that displaces its intellectual predecessors from work that consolidates them. Wu et al. (2019) show that large teams develop existing ideas and small teams disrupt them, a distinction that matters more as team production has become dominant (Wuchty, Jones, and Uzzi 2007; Larivière et al. 2015). Uzzi et al. (2013) show that high-impact papers are usually conventional in their combinations of prior work, with novelty entering through a minority of atypical combinations; Leahey, Lee, and Funk (2023) show that different types of novelty have different consequences for whether work disrupts or consolidates. The amount of knowledge produced and its kind are therefore separable and can be measured apart, which is why we study science rather than another domain of knowledge work.

Based on this discussion, we advance two hypotheses.

***H1.*** *Unexplained pay dispersion is positively associated with the dimensions of output that are observable at the time of evaluation: publication counts and citations.*

***H2.*** *Unexplained pay dispersion is negatively associated with the dimensions that are difficult to evaluate at the time of evaluation, disruption and novelty, and positively associated with conventionality, the extent to which work recombines what the field already holds.*

Three scope conditions bound the argument. First, evaluation must be peer-relative and frame-dependent, so that the signals available to the principal measure conformity to prevailing standards. Second, the lag between frame-revising work and recognition of its value must be longer than the interval at which rewards are set. Third, pay dispersion must be salient enough to shape effort or employer choice. These conditions hold in academic science, and can plausibly extend to corporate research, design, and other knowledge-intensive professions. Where output is judged against a stable, external, and quickly observed criterion, the classical accounts should apply without this compositional qualification.

Two features of the scientific reward system may sharpen the prediction. First, recognition is cumulative (Merton 1968, 1988; Cole and Cole 1973; Zuckerman 1977; Dannefer 2003), so small differences in early legibility are likely to grow into large differences in resources through the same frame-relative signals a tournament uses. Second, the institutions that allocate those resources are themselves conservative. Prior research shows that evaluators give lower scores to highly novel proposals (Boudreau et al. 2016), novel papers reach high-status venues less readily (Wang et al. 2017), and applicants just above a funding threshold accumulate substantially more later resources than those just below (Bol, de Vaan, and van de Rijt 2018).

## 3. Data and methods

### 3.1. Data

To test these hypotheses, we scraped the public administrative payroll records of 93 U.S. public universities, between fiscal years 2004 and 2026, using an AI-supported data collection pipeline (see Appendix A in the online supplement). Public-sector transparency statutes in most states require universities to disclose employee compensation, and the resulting files, released in heterogeneous formats through university offices, state comptrollers, and public-records portals, constitute an underused census of the public academic labor market. We restrict the sample to tenure-track faculty and to departments whose mean yearly tenure-line headcount is between 5 and 200. The upper bound excludes a small number of units, such as medical schools reported as a single department. Each person-year record carries base pay, academic rank, contract basis (nine- or twelve-month), department, and institution. Table A2 in Appendix A reports coverage by institution.

The collection itself departs from prior practice. The payroll records of the 101 universities whose files yield tenure-line records (Appendix A.3) are released in as many formats, and assembling two decades of them by hand is the reason the salary side of the science-of-science literature has remained thin. We delegated acquisition, format normalization, department-name harmonization, and rank coding to a large-language-model coding agent (Claude Code), which composed and executed the several hundred site-specific acquisition and parsing routines the corpus required, under human specification and with every step audited by the author. Appendix A documents the audits run on the data. We checked data quality through manual review and online fact-checking. After cleaning, we also audited institutional coverage and row counts against the original sources, re-counted live source records for a random sample of institutions, and verified the provenance of department and rank codes.

We link each faculty member to their publication record in OpenAlex (Priem, Piwowar, and Orr 2022), an open bibliographic index descended from the Microsoft Academic Graph (Wang et al. 2020), through a name-and-institution matching procedure described in Appendix A. Our unit of analysis is the department-person pair: each scholar has a unique record within each related department, spanning the years the person appears on that department's roster. Department strings are taken from the payroll sources after normalization. We collapse to the person-department records rather than a person-year panel or department-year panel, because coverage windows differ across institutions, so a balanced panel would either discard most departments or confound calendar time with which institutions are observed. The final analytic sample includes 124,056 unique records, covering 105,050 individuals in 4,839 departments at 93 universities.

### 3.2. Measuring salary dispersion beyond department pay ladder

For each department in the data, we first estimate a salary model to remove pay dispersion attributable to the department's formal ladder:

$$\log(\mathrm{salary}_{it}) = \alpha_d + \boldsymbol{\beta}'_d[\mathrm{rank}_{it}, \mathrm{contract\ basis}_{it}, \mathrm{career\ age}_{it}, \mathrm{career\ age}^2_{it}] + \gamma_{dt} + \varepsilon_{it},$$

where *i* indexes individuals, *t* fiscal years, and *d* departments. The subscript on β marks the estimator's central feature: every department receives its own coefficient vector, so the ladder is fitted on the department's own terms. The intercept α absorbs everything constant within a department, including its university and its field. Year effects γdt are estimated department by department, so that an across-the-board raise registers as a shift in the department's mean rather than as inequality.

The residual εit is what the department's own ladder cannot account for: the gap between what a person earns and what their rank, contract basis, and seniority predict they would earn in that department in that year. From this residual, we take the Gini. We add back the department-year mean of log pay and exponentiate the result, so the index is a Gini in dollars. This index is therefore counted as dispersion the ladder does not explain.

Estimating the model department by department, rather than pooling observations with department fixed effects, is a substantive choice: it treats each department as having its own distributional culture. The median department equation explains 56.5 percent of the variance in its members' log pay, with an adjusted R2 of 0.479.

Two features of the estimator matter. First, each department-level regression uses up to seven parameters (an intercept and six slopes), plus T−1 year effects. In small departments, the model can therefore interpolate its own members, mechanically shrinking the residuals. Because department size is included as a control in every second-stage model, an uncorrected residual Gini would carry this estimator behavior into the inequality measure. We therefore use the leverage-corrected residual, $\tilde{e}_i = \frac{e_i}{\sqrt{1-h_{ii}}}$, whose second moment is $\sigma^2$ under a correctly specified model with homoskedastic, uncorrelated errors. When departments are grouped by person-years per estimated parameter, the ratio of the department-specific residual Gini to the same department's pooled-equation Gini ranges from 0.68 to 0.95 before correction, and from 0.92 to 0.97 after correction (Appendix Figure A1). Second, 12.8 percent of departments have fewer than five residual degrees of freedom and cannot be estimated on their own terms. For these departments, constructing a residual would require borrowing another department's reward function, precisely the assumption this design is meant to avoid. The analytic sample is therefore limited to the 4,839 departments that can fit their own equation.[1,2]

Figure 1 shows what the residualization accomplishes. Across the analytic departments the median raw pay Gini is 0.153; after each department's own ladder is removed it is 0.108, a reduction of 29 percent. On the variance scale the same first stage explains more, 56 percent of the variance in log pay at the median department, leaving 44 percent unexplained (Table 1). The two figures are the same decomposition read on two scales, and a Gini is far less sensitive than a variance to removing a few large, well-predicted gaps. We use the Gini because it is the quantity this literature regresses on performance, and we report the variance share alongside it. The residual and raw Ginis correlate at 0.86 across records, a high correlation that still leaves roughly a quarter of the variance, which is where our identifying variation resides.

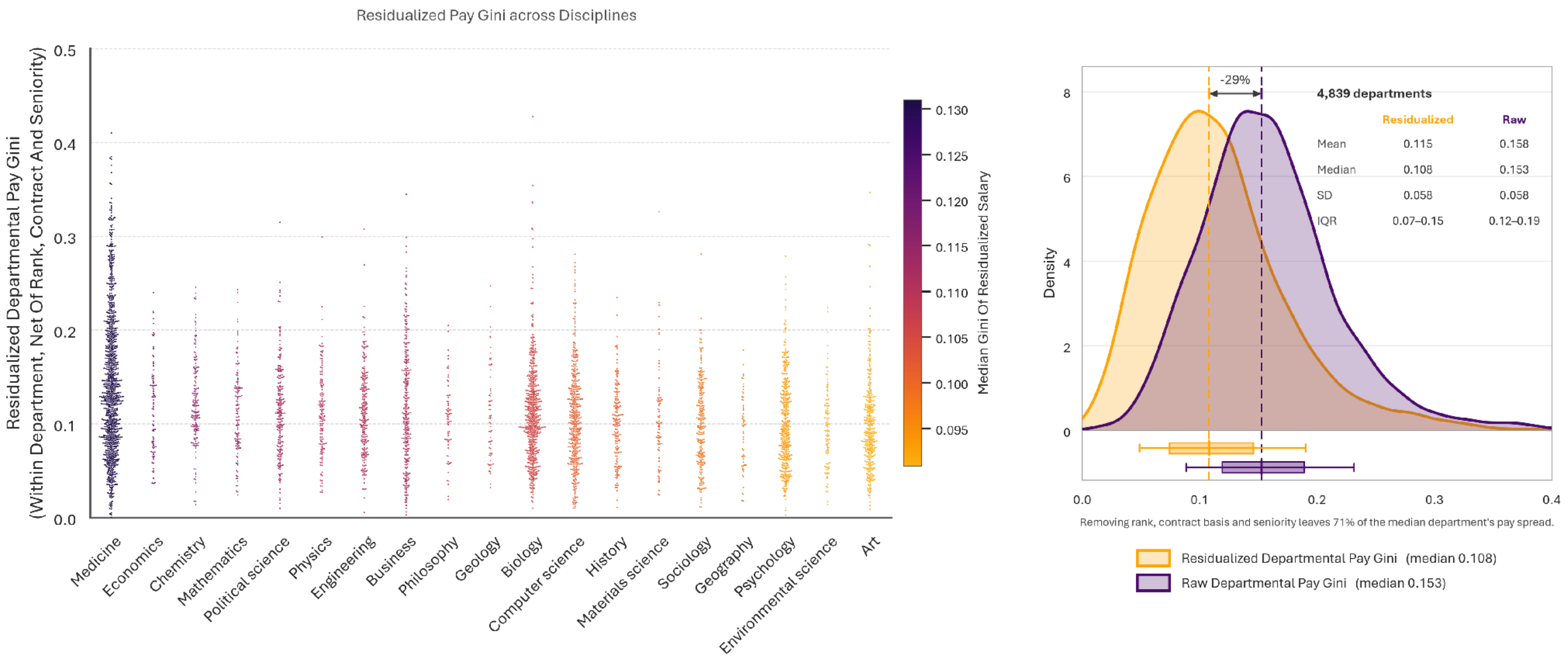

Figure 1. Residualized Pay Dispersion Within Departments at U.S. Public Universities, by Discipline

The same first stage yields a second quantity: the Gini of fitted pay, what each person would earn if paid exactly what the ladder predicts, which measures the dispersion the ladder justifies. We call this the ladder Gini. It correlates with the residualized Gini at 0.27 across individual records and 0.32 across departments, low enough for both to enter one model. Section 4.3 compares the ladder Gini and the raw Gini of observed pay with the residualized Gini.

### 3.3. Measuring the composition of knowledge work

For each scholar in our records, we identify every publication the person produced during the years they sat in that department and compute five outcomes over that set of works, aggregating over works to individual level. The analysis uses five dependent variables, as defined below.

**Disruption.** We measure disruption using three-year Funk–Owen-Smith CD index (CD3). Widely used in research on scientific and technological innovation, this measure captures whether subsequent works cite a focal paper in place of, or alongside, the prior literature on which it builds (Funk and Owen-Smith 2017; Wu et al. 2019; Park, Leahey, and Funk 2023). For each focal paper f, the index distinguishes works in its citation window that cite either f or at least one of its references. CD3 is defined as:

$$CD_3(f) = \frac{n_i - n_j}{n_i + n_j + n_k}$$

where $n_i$ counts works citing $f$ but none of its references; $n_j$ counts works citing both $f$ and at least one of its references; and $n_k$ counts works citing at least one of its references but not $f$. Positive values indicate a more disruptive citation pattern, whereas negative values indicate a more consolidating pattern. The index ranges from $-1$, when all works in the relevant citation set cite both the focal paper and its antecedents, to $+1$, when all cite the focal paper without citing its antecedents. We computed CD3 based on the June 2026 OpenAlex snapshot, for all scientific publications published in 2023 or earlier.

**Novelty versus Conventionality.** Following Uzzi et al. (2013), we assess whether a paper combines sources that are cited together less or more often than expected by chance. For each pair of journals or conference series represented in its references, we calculate:

$$z_{ab} = \frac{O_{ab} - \mu_{ab}}{\sigma_{ab}}$$

where $O_{ab}$ is the observed frequency with which venues *a* and *b* are cited together, and $\mu_{ab}$ and $\sigma_{ab}$ are the mean and standard deviation of that frequency across randomized citation networks preserving citation counts and year structure. Negative values indicate atypical pairings; positive values indicate more conventional combinations. For each paper *f,* we summarize its distribution of pairwise scores, $Z_f$, using:

$$Novelty_f = -Q_{0.10} Z_f, \ Conventionality_f = Q_{0.50} Z_f$$

where $Q_{0.10}$ and $Q_{0.50}$ denote the tenth percentile and median, respectively. Novelty captures the most unusual combinations among a paper's references; we reverse its sign so that higher values indicate greater novelty. Conventionality captures how familiar its typical combination is. Note that these two values do not often go in the same direction: a paper can score highly on both by drawing mainly on established connections while introducing a few unusual ones: a pattern associated with high scientific impact (Uzzi et al. 2013). Details of the randomization procedure are provided in Uzzi et al. (2013).

**Citations.** Citations are counted as the number of forward citations received in the publication year and the two following years; as with CD3, only works published through 2023 enter, so every count covers a complete window. Novelty and conventionality likewise use works published through 2023, so the four work-level outcomes share the same publication-year cutoff, and scholars whose department windows begin in 2024 or later contribute only to productivity.

**Productivity.** Productivity is calculated as the average publications per year, measuring the scholar's publication speed. It is defined for every scholar with an OpenAlex identity, including those with no work in the window, who enter with a productivity of zero; the other four outcomes are defined only for scholars with at least one work carrying the references each measure requires (Table 1). The first four are standardized as z-score against the full OpenAlex distribution of works published in the same year before aggregation. All five are rank-normalized before estimation, because the underlying distributions have tails in which a handful of works would otherwise dominate the estimate.

Although we use the CD index and atypicality as focal measures, both have known limitations. The CD index is sensitive to the length of the forward citation window, can be biased over time by changes in reference lists and citation volume, and varies across specifications (Bornmann and Tekles 2019; Leibel and Bornmann 2024; Petersen, Arroyave, and Pammolli 2024). The Uzzi measures face a related concern: an unusual pairing of cited journals is not the same as a novel idea (Bornmann et al. 2019). Two of these concerns involve comparability across publication years. We address differences across publication cohorts by standardizing each work against all OpenAlex works published in the same year. This adjusts for cohort-specific means and scales, but does not eliminate potential bias related to citation windows or database coverage. Because the analysis compares scholars within publication years rather than estimating a long-run trend in disruption, it is less exposed to recent concerns about secular changes in the CD index (Holst et al. 2026; Park, Leahey, and Funk 2026).

The construct-validity concern remains. Our design never interprets their absolute levels. All estimates compare departments within fields and universities, so a measurement bias would threaten the results only if it were correlated with a department's unexplained pay dispersion. Our claims are therefore about how measured features of the citation record covary with an organizational variable.

### 3.4. Controls and estimation

All second-stage models take the form

$$Y_{ij} = \beta\,\mathrm{Gini}_j + \mathbf{X}'_{ij}\gamma + \alpha_{u(j)} + \delta_{f(j)} + \varepsilon_{ij}$$

where *i* indexes a person, *j* a department, *u* a university, and *f* one of the 26 OpenAlex fields. Controls include the scholar's career age in the observed year closest to the middle of their time in the department (career age is measured from one first-publication year per scholar), its square, inferred gender and race, the highest rank the person reached in that department, the department's share female and share minority, log department size, the department's years in the data, the person's first year in the department, and the number of years they spent there. Gender and race are inferred from names, with supplementary photo-based classification, and used only as coarse controls and matching strata; their coefficients are not interpreted (Lockhart, King, and Munsch 2023).

We cluster standard errors by university. Every specification is reported with university and field fixed effects, and again with university-by-field fixed effects. The latter compares departments only with other departments in the same broad field at the same university. To assess sensitivity to first-stage estimation error in the Gini, we use a within-department wild bootstrap that redraws each scholar's residuals jointly across years and add the resulting variance component to the cluster-robust second-stage variance. Factors estimated for the Table 2 models are reused for the corresponding estimated-Gini coefficients in Table 3, Figure 3 except the matched sample, and Appendix Tables B1, B5 and B6. The raw Gini of observed pay in Table 3 retains its cluster-robust standard error. In Table 2, the adjustment increases standard errors by 0.6 to 2.3 percent and leaves the substantive conclusions unchanged.

Four checks bear on the construction of the measure: the full panel including departments too small for their own equation, the uncorrected residual, a first stage with the h-index and grant income returned, and coarsened exact matching (Iacus, King, and Porro 2011, 2012) on content cluster, career-age band, gender, race, broad area, Times Higher Education status, department size quintile, and years on record. Two further designs address the team-composition and selection accounts described below. A final design uses the retirement and the arrival of a department's top-paid senior member as within-department tests. The results of these robustness tests are reported in Figure 3 and in Appendix B.

## 4. Results

### 4.1. The shape of unexplained dispersion

Table 1 reports descriptive statistics at both levels. Across departments the residual pay Gini has a median of 0.108 and an interquartile range of 0.074 to 0.146. Weighted by department-individual records, the unit the models use, the median is 0.123 and the interquartile range 0.089 to 0.159, because larger departments are both more numerous in the number of members and somewhat more unequal.

In terms a faculty member would recognize, a Gini of 0.063, the tenth percentile across records, implies that two randomly selected colleagues differ in pay by approximately 13 percent of the departmental mean once the ladder has been accounted for. A Gini of 0.209, the ninetieth percentile, implies a difference of approximately 42 percent. At the mean salary of $127,216 these are the $16,000 and $53,200 of the opening paragraph.

Table 1. Descriptive Statistics

| | **Mean** | **SD** | **P10** | **Median** | **P90** | **N** |
|---|---|---|---|---|---|---|
| *Panel A. Departments (N = 4,839)* | | | | | | |
| Residualized pay Gini | 0.115 | 0.058 | 0.048 | 0.108 | 0.190 | |
| Ladder pay Gini | 0.108 | 0.046 | 0.055 | 0.103 | 0.166 | |
| Raw pay Gini | 0.158 | 0.058 | 0.088 | 0.153 | 0.231 | |
| Share of pay variance unexplained | 0.447 | 0.227 | 0.143 | 0.435 | 0.763 | |
| Department equation R-squared | 0.553 | 0.227 | 0.237 | 0.565 | 0.857 | |
| Department equation adjusted R-squared | 0.460 | 0.268 | 0.093 | 0.479 | 0.808 | |
| Residual degrees of freedom | 102.3 | 175.6 | 9.000 | 42.0 | 252.2 | |
| Department size (mean yearly) | 17.5 | 15.7 | 5.714 | 13.0 | 32.4 | |
| Share female | 0.365 | 0.187 | 0.143 | 0.345 | 0.619 | |
| Share minority | 0.275 | 0.173 | 0.071 | 0.250 | 0.500 | |
| Years on record | 6.625 | 5.470 | 1.000 | 5.000 | 15.0 | |
| *Panel B. Department-person records (N = 124,056)* | | | | | | |
| Career age | 18.1 | 11.7 | 4.000 | 16.0 | 36.0 | 124,056 |
| h-index | 7.075 | 11.7 | 0.000 | 2.000 | 22.0 | 124,056 |
| Years in the department | 4.411 | 4.114 | 1.000 | 3.000 | 11.0 | 124,056 |
| Salary (USD) | 127,216 | 70,792 | 63,659 | 110,127 | 211,318 | 124,056 |
| Pay premium (log points) | -0.020 | 0.290 | -0.324 | -0.001 | 0.274 | 123,894 |
| Paper Citations (C3) | 0.437 | 1.078 | -0.073 | 0.214 | 1.040 | 66,033 |
| Disruption (CD3) | -0.046 | 0.483 | -0.143 | -0.082 | -0.026 | 61,637 |
| Novelty | 0.024 | 0.265 | -0.046 | 0.061 | 0.121 | 60,184 |
| Conventionality | -0.025 | 0.449 | -0.260 | -0.128 | 0.232 | 60,184 |
| Productivity per year | 8.003 | 46.5 | 0.111 | 2.000 | 10.5 | 94,564 |
| Team size (log) | 1.772 | 0.539 | 0.938 | 1.835 | 2.408 | 84,574 |
| Solo-authored share | 0.144 | 0.289 | 0.000 | 0.000 | 0.667 | 85,613 |
| Topic HHI | 0.386 | 0.326 | 0.073 | 0.265 | 1.000 | 85,552 |
| Distinct topics | 18.4 | 76.3 | 1.000 | 5.000 | 28.0 | 85,613 |

Note: Panel A describes the 4,839 departments that fit their own pay equation; Panel B the 124,056 department-person records in those departments. The residualized pay Gini is the Gini of leverage-corrected residuals from each department's own pay equation, recentred on the department-year mean and exponentiated; the ladder Gini is the Gini of fitted pay from the same equation. Disruption, citations, novelty and conventionality are z-scored against all OpenAlex works published in the same year before aggregated to individual level; productivity is publications per year. N differs across outcome rows because not every scholar has a linked publication record with the references needed for each measure.

Figure 1 plots each analytic department as a point. Even within the same broad area, departments range from near-perfect internal equity to residual Ginis above 0.30. In most disciplines, the 90th-percentile department is roughly three to four times as unequal as the 10th-percentile department; across disciplines the ratio ranges from about 2.5 to 5.5. Unexplained dispersion is therefore a property of departments, not simply of the disciplinary categories they belong to. This is why we treat it as an organizational variable and compare departments only within fields, and within universities.

### 4.2. Dispersion and the composition of knowledge work

Table 2 reports our main OLS estimates. We report the main estimates as standardized coefficients, so each entry is the difference in the outcome, in standard deviations, associated with a one-standard-deviation difference in the regressor. The pattern is consistent across all five outcomes and both fixed-effect sets.

Scholars in departments with greater unexplained pay dispersion are more productive and more highly cited. A one-SD increase in unexplained dispersion is associated with a 0.040-SD increase in productivity ($p = 0.015$) and a 0.049-SD increase in citations ($p < 0.001$). Both associations are larger under university-by-field fixed effects, at 0.048 and 0.062, respectively. These magnitudes are about two-thirds to three-quarters of the female–male gap in the same outcomes and models (women's productivity is 0.063 SD and their citations 0.067 SD lower than men's; these are female–male differences in outcome SDs, whereas Table 2 reports the female coefficient fully standardized, per SD of the 0/1 indicator, as −0.030 and −0.031). H1 is therefore supported: both legible dimensions of output are higher where unexplained dispersion is greater.

This pattern is consistent with tournament theory, and contrasts with early survey-based evidence from academic settings, which found that pay dispersion was associated with lower self-reported productivity (Pfeffer and Langton 1993). In our data, unexplained pay dispersion is instead associated with higher output on the most legible dimensions of performance.

Table 2. Departmental Pay Inequality and Scholars' Knowledge Production Style

| | **Disruption (CD3)** | | **Novelty** | | **Conventionality** | | **Citations (C3)** | | **Productivity** | |
|---|---|---|---|---|---|---|---|---|---|---|
| | Univ. + field FE | Univ. × field FE | Univ. + field FE | Univ. × field FE | Univ. + field FE | Univ. × field FE | Univ. + field FE | Univ. × field FE | Univ. + field FE | Univ. × field FE |
| Residualized Pay Gini | -0.029* | -0.035* | -0.025* | -0.023* | 0.036** | 0.043** | 0.049*** | 0.062*** | 0.040* | 0.048* |
| | (0.013) | (0.014) | (0.010) | (0.011) | (0.011) | (0.012) | (0.009) | (0.011) | (0.016) | (0.019) |
| Career Age (Mid During Stay) | -0.036* | -0.034* | 0.022 | 0.023 | 0.012 | 0.010 | 0.091*** | 0.092*** | 0.085*** | 0.083*** |
| | (0.015) | (0.015) | (0.014) | (0.014) | (0.015) | (0.015) | (0.017) | (0.017) | (0.020) | (0.019) |
| Career Age Squared (×100) | 3.914** | 3.710* | -1.731 | -1.856 | -0.402 | -0.360 | -9.809*** | -9.872*** | -9.205*** | -9.070*** |
| | (1.391) | (1.404) | (1.474) | (1.463) | (1.489) | (1.505) | (1.516) | (1.486) | (1.737) | (1.664) |
| Gender: Female (ref: Male) | 0.023*** | 0.023*** | 0.005 | 0.005 | -0.047*** | -0.047*** | -0.031*** | -0.032*** | -0.030*** | -0.031*** |
| | (0.004) | (0.004) | (0.003) | (0.004) | (0.005) | (0.005) | (0.004) | (0.004) | (0.004) | (0.005) |
| Gender: Unknown (ref: Male) | -0.006 | -0.006 | 0.002 | 0.002 | -0.000 | 0.000 | -0.010** | -0.010** | -0.011* | -0.012** |
| | (0.006) | (0.005) | (0.006) | (0.006) | (0.006) | (0.006) | (0.004) | (0.004) | (0.004) | (0.004) |
| Race: Asian (ref: White) | 0.026*** | 0.026*** | -0.058*** | -0.058*** | 0.098*** | 0.097*** | 0.073*** | 0.073*** | 0.256*** | 0.256*** |
| | (0.004) | (0.004) | (0.005) | (0.005) | (0.006) | (0.006) | (0.007) | (0.008) | (0.009) | (0.009) |
| Race: Hispanic (ref: White) | 0.012** | 0.011* | -0.005 | -0.005 | 0.001 | 0.001 | -0.024*** | -0.024*** | -0.020*** | -0.019*** |
| | (0.004) | (0.004) | (0.004) | (0.004) | (0.005) | (0.005) | (0.005) | (0.005) | (0.005) | (0.005) |
| Race: Non-Hispanic Black (ref: White) | 0.006 | 0.006 | -0.002 | -0.002 | -0.009+ | -0.007 | -0.013** | -0.012** | -0.009** | -0.009** |
| | (0.004) | (0.004) | (0.005) | (0.005) | (0.005) | (0.005) | (0.004) | (0.004) | (0.003) | (0.003) |
| Highest Rank | 0.023*** | 0.023*** | -0.011+ | -0.012* | 0.004 | 0.004 | -0.012 | -0.014 | 0.078*** | 0.077*** |
| | (0.005) | (0.005) | (0.006) | (0.006) | (0.006) | (0.006) | (0.008) | (0.008) | (0.013) | (0.013) |
| Department Share Female | 0.120*** | 0.126*** | 0.024* | 0.029** | -0.162*** | -0.169*** | -0.073*** | -0.068*** | -0.002 | 0.002 |
| | (0.017) | (0.017) | (0.009) | (0.010) | (0.011) | (0.011) | (0.009) | (0.009) | (0.010) | (0.010) |
| Department Share Minority | 0.022** | 0.016* | 0.024** | 0.030** | -0.025* | -0.028* | 0.017+ | 0.031** | 0.025+ | 0.034** |
| | (0.008) | (0.008) | (0.008) | (0.010) | (0.010) | (0.012) | (0.010) | (0.009) | (0.013) | (0.012) |
| Log(Department Size) | -0.013 | -0.007 | -0.009 | -0.008 | 0.014 | 0.015 | -0.003 | -0.009 | -0.032 | -0.028 |
| | (0.010) | (0.011) | (0.010) | (0.014) | (0.009) | (0.010) | (0.011) | (0.012) | (0.021) | (0.022) |
| Years on Record | 0.007 | 0.008 | 0.020* | 0.019* | -0.016 | -0.010 | -0.011 | 0.005 | 0.008 | 0.023 |
| | (0.017) | (0.018) | (0.009) | (0.009) | (0.010) | (0.013) | (0.010) | (0.011) | (0.035) | (0.031) |
| Years in Department | -0.039** | -0.041** | -0.208*** | -0.207*** | 0.195*** | 0.193*** | 0.033** | 0.032** | 0.054** | 0.051*** |
| | (0.012) | (0.013) | (0.011) | (0.010) | (0.008) | (0.008) | (0.011) | (0.010) | (0.016) | (0.013) |
| First Year in Department | -0.171*** | -0.170*** | -0.335*** | -0.334*** | 0.257*** | 0.258*** | -0.103*** | -0.107*** | 0.018 | 0.021 |
| | (0.013) | (0.014) | (0.011) | (0.011) | (0.010) | (0.010) | (0.014) | (0.014) | (0.026) | (0.023) |

| | Disruption (CD3) | | Novelty | | Conventionality | | Citations (C3) | | Productivity | |
|---|---|---|---|---|---|---|---|---|---|---|
| | Univ. + field FE | Univ. × field FE | Univ. + field FE | Univ. × field FE | Univ. + field FE | Univ. × field FE | Univ. + field FE | Univ. × field FE | Univ. + field FE | Univ. × field FE |
| University FE | YES | | YES | | YES | | YES | | YES | |
| Field Fixed Effects | YES | | YES | | YES | | YES | | YES | |
| Univ. × field FE | | YES | | YES | | YES | | YES | | YES |
| Adjusted within R-squared | 0.031 | 0.030 | 0.069 | 0.067 | 0.080 | 0.079 | 0.031 | 0.032 | 0.084 | 0.085 |
| Adjusted R-squared | 0.106 | 0.118 | 0.111 | 0.120 | 0.152 | 0.165 | 0.186 | 0.205 | 0.168 | 0.185 |
| Number of Observations | 61,605 | 61,591 | 60,154 | 60,138 | 60,154 | 60,138 | 65,996 | 65,991 | 94,488 | 94,467 |

Note: Standard errors in parentheses. + $p < 0.1$, * $p < 0.05$, ** $p < 0.01$, *** $p < 0.001$. All tests are two-tailed. Dependent variables are rank-normalized. Univ. denotes university; FE denotes fixed effects. Reference categories are male for gender and non-Hispanic white for race. Coefficients are standardized, including those of binary indicators (per SD of the indicator).

However, these same departments produce work that less often displaces its predecessors and recombines prior literature more conventionally. A one-SD increase in unexplained dispersion is associated with a 0.029-SD decrease in disruption (p = 0.029), a 0.025-SD decrease in novelty (p = 0.017), and a 0.036-SD increase in conventionality (p = 0.001). Under university-by-field fixed effects, the disruption and conventionality associations are stronger, at -0.035 and 0.043 respectively, while the novelty association is essentially unchanged at -0.023 (p = 0.045). Therefore, H2 is supported across all three compositional measures. This pattern is consistent with our argument that highly novel work, being harder to evaluate at the moment it is produced, is less likely to be rewarded by tournaments: scholars in departments with greater unexplained pay dispersion produce work that is, on average, more conventional, less disruptive, and less novel.

The magnitudes are modest by the standards of the individual-level literature, but unexceptional for a department-level covariate. We emphasize that this is an association between an organizational property and the average character of a department's output, not a causal model of why any particular scholar's work is disruptive. Its interpretive weight rests on the sign pattern across all five outcomes: a pattern that is unlikely under any account in which dispersion is uniformly associated with higher or lower performance, and one that reappears under stricter fixed effects that would ordinarily attenuate a spurious department-level association.

### 4.3. Raw, ladder-explained, and residualized inequality

Readers might reasonably ask what the results look like when salary differences associated with rank are not conditioned away. Academic careers are rewarded by promotion as much as by pay, and a department whose ladder is steep may be one that rewards early-career science through rank rather than through the salary. We therefore compute, for every department-year and from the same first stage, three measures of pay dispersion: the raw Gini of observed pay; the ladder Gini of pay fitted only from rank, contract basis and seniority; and the residualized Gini used so far. Across departments, the residualized Gini correlates 0.81 with the raw Gini, and 0.32 with the ladder Gini.

Table 3 reports the main coefficients for each measure, when we exchange them into the Table 2 specification. The sign pattern is identical for every measure. For disruption, citations, and productivity, Raw Gini yields the largest estimates. Rank adjustment therefore does not alter the direction of these associations. For conventionality, the coefficient on Raw Gini is smaller than that on residualized Gini, and all three coefficients keep the same sign, although the ladder estimate is only marginally significant ($p = 0.065$). The ladder Gini is not significant for novelty ($p = 0.43$) and only marginally significant for citations ($p = 0.076$), showing that pay disparities arising from the shared ladder are less associated with the research style outcomes than unexplained inequality is.

Residualization therefore matters most for novelty and conventionality, which are tied most strongly to the dispersion the ladder does not explain; for the other outcomes it changes the size of the associations but not their direction.

Table 3. Alternative Constructions of Departmental Pay Inequality

| **Outcome** | **Raw Gini** | **Ladder Gini** | **Residualized Gini** |
|---|---|---|---|
| Disruption | -0.043***<br>(0.010) | -0.037***<br>(0.008) | -0.029*<br>(0.013) |
| Novelty | -0.015*<br>(0.007) | -0.007<br>(0.009) | -0.025*<br>(0.010) |
| Conventionality | 0.027*<br>(0.013) | 0.018+<br>(0.009) | 0.036**<br>(0.011) |
| Citations | 0.060**<br>(0.018) | 0.042+<br>(0.023) | 0.049***<br>(0.009) |
| Productivity | 0.069**<br>(0.023) | 0.057*<br>(0.026) | 0.040*<br>(0.016) |

*Note: Standardized coefficients (SD of the outcome per SD of the measure), university + field fixed effects, Table 2 controls, standard errors clustered on university in parentheses. Each measure is computed from the department's own first stage: the raw Gini of observed pay; the ladder Gini of pay fitted from rank, contract basis and seniority; the residualized Gini of the leverage-corrected residual. Each column enters one measure at a time. + $p < 0.1$, * $p < 0.05$, ** $p < 0.01$, *** $p < 0.001$.*

We also discuss the associations between research styles and promotion (instead of salary) in Appendix B, Table B3.

## 4.4. What departments pay for above their ladder

The first stage yields, for every person-year, the individual's residual pay: what that person is paid above or below what their department's formal ladder predicts. This is the micro-level counterpart of the department-level measure. Asking what it correlates with tells us what unexplained pay rewards.

Figure 2 examines the relationship between research style and the pay premium not explained by the department's formal pay ladder. We divide each of six research-style measures into deciles and estimate the pay premium for each decile relative to the bottom decile, net of the full control set. To address the possibility that some research styles are correlated with performance measures, such as higher productivity in large teams, we estimate a second set of department-level pay equations that also control for merit-based indicators, including the h-index and funding attainment. We then calculate a new residual premium from these models. Figure 2 reports results for both sets of payment premium, with similar patterns.

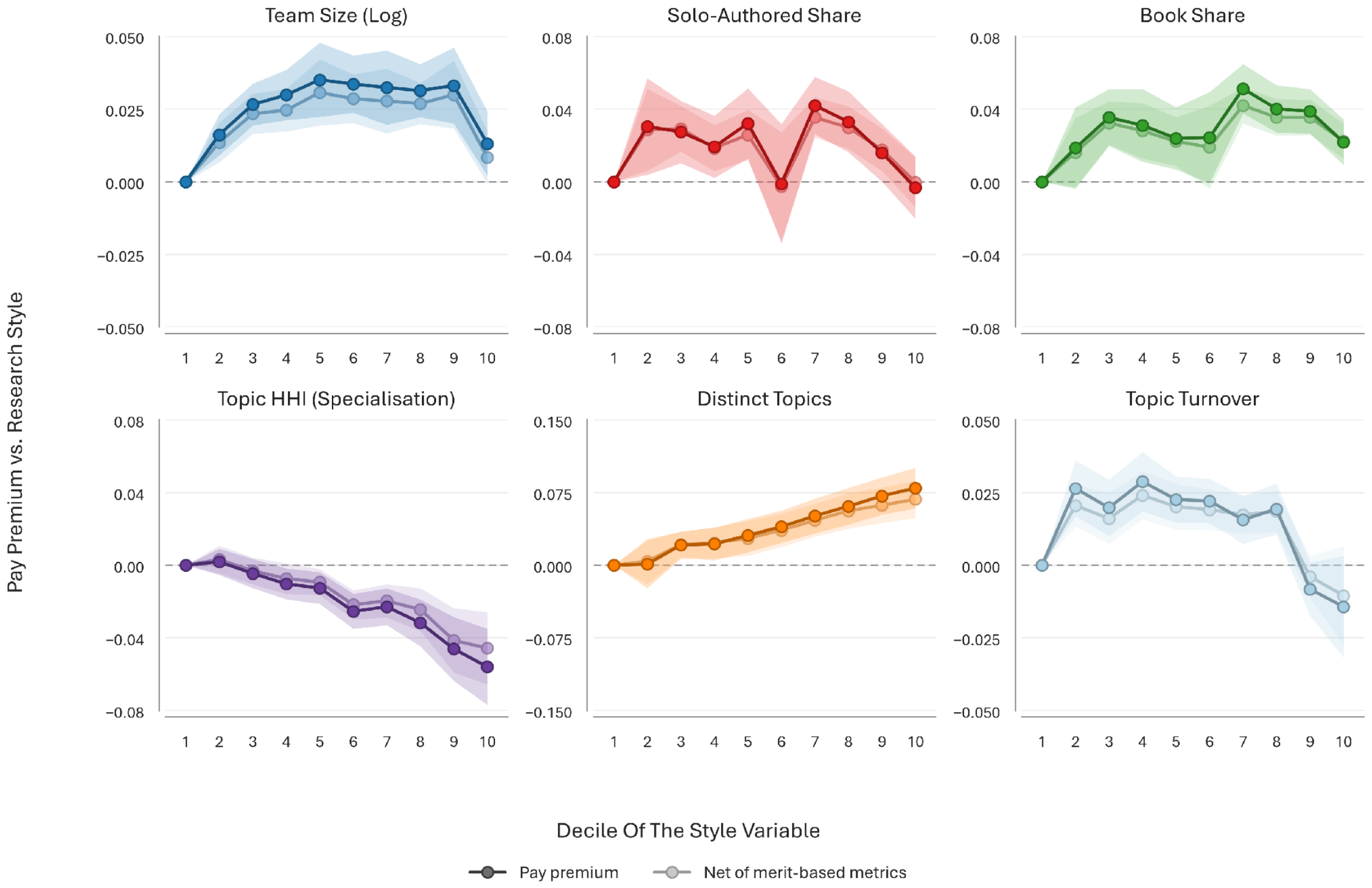

Figure 2. Research Style and Payment Premium in Academic Settings

The profile corresponds to a recognizable type: departments reward the scholar who works across many topics and does not specialize narrowly. The premium rises most steeply with the number of distinct topics a person works on (top decile 0.080 log points, $p < 0.001$) and falls with topic concentration (−0.056, $p < 0.001$); it rises only modestly with team size (0.013, $p = 0.022$) and with the share of output that is books (0.022, $p = 0.001$). Solo authorship and topic turnover are priced non-monotonically: the premium differs significantly across their deciles (joint tests of the nine decile coefficients, $p < 0.001$ for both), but it is higher in the middle deciles than at either end, so the top decile does not differ from the bottom one ($p = 0.71$ and $p = 0.11$).

Therefore, breadth commands a premium; depth does not, and solo authorship and topical mobility are rewarded at moderate levels rather than at either extreme. Beyond the ladder, departments appear to pay for the profile of the productive, well-connected, broadly engaged researcher, which is the profile frame-conforming knowledge production selects for (Wu et al. 2019).

The department-level counterpart, regressing six style measures on the residual Gini with the Table 2 controls, agrees. A one-SD difference in unexplained dispersion is associated with larger teams (0.115 standard deviations, $p < 0.001$), less solo authorship (−0.054, $p < 0.001$), marginally lower topic concentration (−0.020, $p = 0.060$), and more distinct topics (0.025, $p = 0.039$).

### 4.5. Robustness Tests

To test the robustness of our findings, we compare the main estimates with results from alternative designs, all reported in Figure 3.

First, we estimate the pay ladder on the full panel, allowing departments too small to fit their own equation to receive a pooled residual.

Second, we add merit-based measures, including h-index and grants, to each department's pay equation, so that the residual is measured net of the performance criteria a department could most defensibly reward.

Third, we drop the top 10 percent of scholars in each department by team size to rule out the possibility that large teams alone drive the patterns we observe.

Finally, we use a matching strategy. Applying coarsened exact matching (CEM) retains 97,161 of 124,056 records across 4,312 strata, restricting comparisons to the same coarsened strata for department size, career age, gender, race, broad disciplinary area, university ranking status, years on record, and research topic. To capture intellectual proximity, we first use k-means clustering in semantic space to identify groups of scholars working on similar topics, and then match scholars within the same cluster.

For continuous-Gini regressions, coefficients express the outcome in standard deviations per one-SD difference in Gini. The CEM estimate is instead a high- versus low-Gini group contrast, divided by the unweighted group-mean Gini difference in the full eligible panel, then multiplied by that panel's Gini SD and divided by the outcome SD in the model's complete-case sample. This linear rescaling aids visual comparison, but does not estimate a continuous-Gini slope.

As Figure 3 shows, the results broadly confirm the robustness of our findings. Although CEM matching leaves two estimates without significance (disruption, $p = 0.41$; productivity, $p = 0.23$), the overall pattern remains intact: novelty ($p < 0.001$), conventionality ($p = 0.021$) and citations ($p = 0.003$) keep their signs and significance in the matched sample.

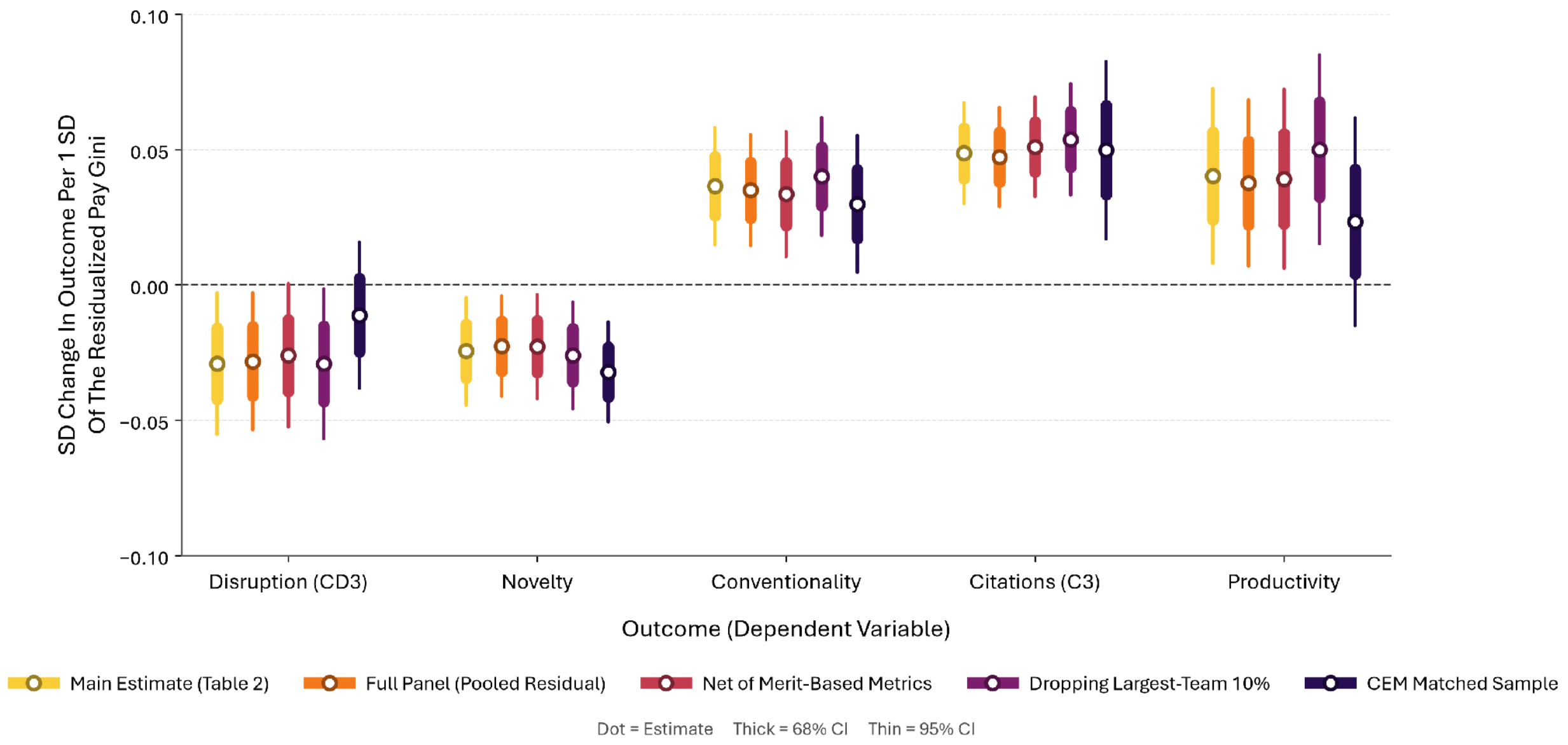

Figure 3. Robustness Tests Across Fixed Effects, Sample Restrictions, and Matching Strategies

### 4.6. A within-department test: the retirement and arrival of a top-paid senior member

The evidence so far compares departments. Only about one-quarter of the yearly Gini's variance lies within departments, and the yearly Gini contains estimation noise of its own, so within-department null results would provide only weak evidence of absence. They do, however, motivate a design that isolates discrete changes in a department's pay dispersion: the retirement and the arrival of a top-paid senior member, which should move dispersion in opposite directions.

The change we use is the retirement of one of the department's senior and highly paid members. We focus on senior members because their exits are less likely to reflect current professional performance and more likely to occur for age-related reasons, and therefore less likely to reflect the department's own research trajectory. We also focus on highly paid members because their departure is the exit most likely to change the department's residual pay dispersion. Specifically, we identify individuals who were in the top decile of their department's pay-premium distribution for two consecutive years, whose final year on the university roster occurs while the roster continues, who do not reappear at another sampled university, and whose career age at exit is at least 30 years since first publication. Together, these criteria identify exits that are most consistent with patterns of retirement. We use the department's first such event as the shock and require the department to be observed for at least two years before and after the event. This standard leaves us with 307 events. As a mirror test, we also identify arrivals: individuals whose first year at a university falls after the university enters the data, who are in the top decile of the department's pay-premium distribution in their first two years there, and whose career age at arrival is at least 30 years. The same department-level rules leave 159 arrival events. For each design, the control group consists of departments that had no comparable event from four years before to five years after the event but did have at least one comparable senior member, who was highly paid and stayed through the year after the event. Within each field and size tertile of head count (or within the field when the tercile holds fewer than three candidates), we select up to twenty nearest control departments based on log size, the two prior Gini values, and mean career age, and assign each of the K selected control departments weight 1/K. These department-level weights sum to one within a matched set; after expansion to incumbent person-years, treated and control groups need not have equal total weight.[3]

Based on these matched samples, we estimate a stacked difference-in-differences by event cohort, separately for retirements and arrivals, with standard errors clustered at the university level. Coefficient tests and pointwise 95 percent confidence intervals use the t distribution with G−1 degrees of freedom, where G is the model-specific number of university clusters. The pre-event test uses F(1, G−1), equivalent to the squared t test for the single pre-event coefficient. Outcomes are measured among incumbents, the people on the roster the year before the event other than the leaver, so that neither the newcomer nor any other new hire enters the outcome, and they are estimated on the incumbents' own person-years with cohort-by-department-by-person and cohort-by-year fixed effects, so that each incumbent is compared with themselves before and after the event. Event time is grouped into two-year periods, from four years before to five years after the event, with the two years before the event as the reference. Incumbent outcomes are rank-normalized. The residual Gini remains in Gini units.

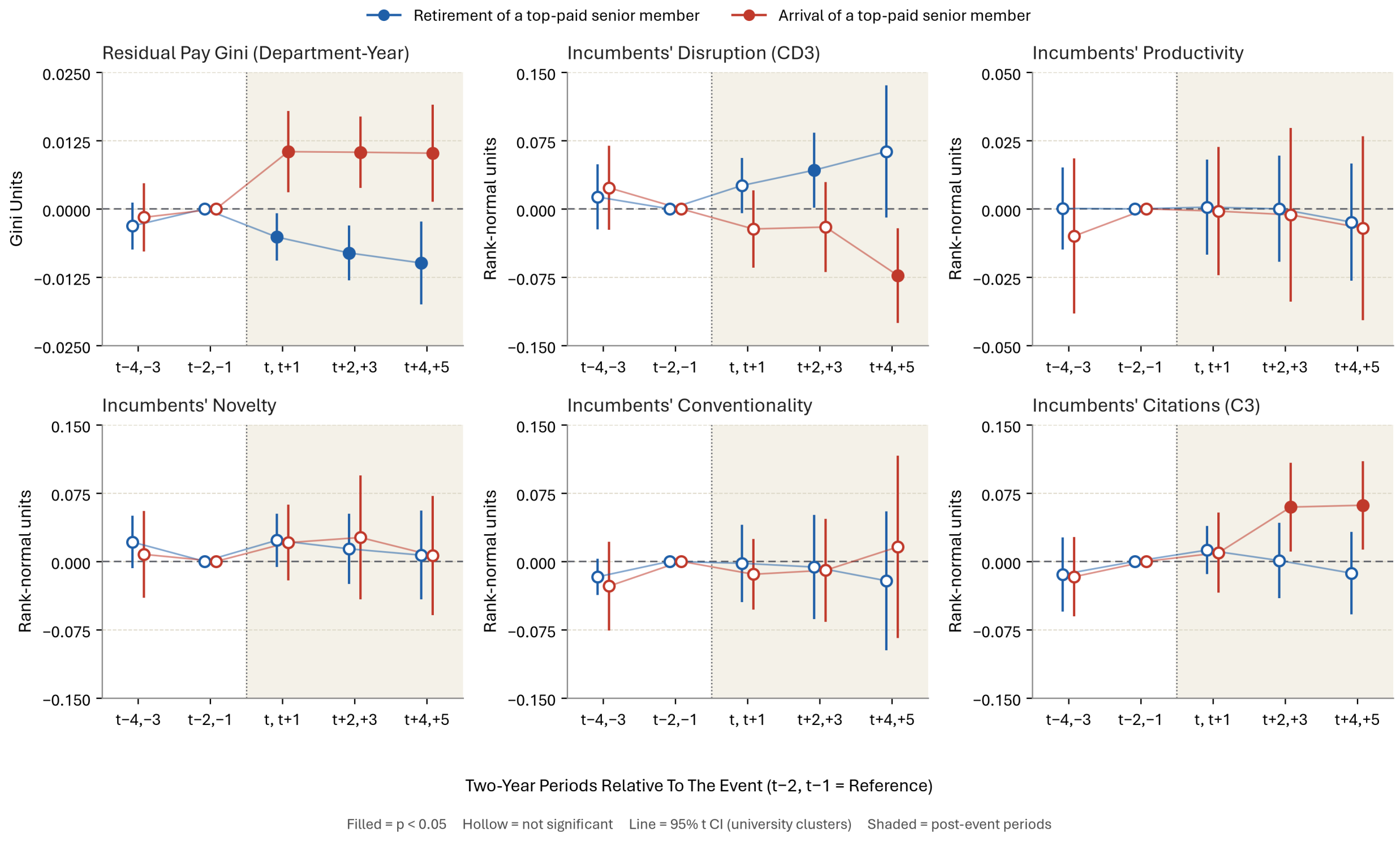

Figure 4. Retirement and Arrival of Top-Paid Senior Faculty and Research Styles of Incumbents

Note: Intervals use t(G−1) critical values with university-clustered standard errors; research outcomes are in rank-normal units. Tests and intervals are not adjusted for multiple comparisons.

As Figure 4 shows, the two events move the department's unexplained pay dispersion in opposite directions. After a retirement, the residual Gini falls in every two-year period, from −0.005 Gini units in the event period to −0.010 four to five years later, with no statistically detectable pre-event difference (p = 0.13). Pooled over the periods after the event period, the fall is 0.007 Gini units (p = 0.014), about 7 percent of the median department-year Gini and one-quarter of its within-department standard deviation. After an arrival, the residual Gini rises by about 0.010 Gini units in every period and stays at that level (pooled 0.011, p = 0.002; pre-event p = 0.61), about 10 percent of the median and one-third of the within-department standard deviation. Both events therefore change the department's unexplained dispersion, if modestly.

The output of the colleagues who remain shifts in the opposite direction. After a retirement, their disruption rises 0.026 in the event period (p = 0.08), by 0.043 in the following two years (p = 0.04), and by 0.063 four to five years later (p = 0.08); pooled over the periods after the event period, it rises 0.045 (p = 0.085). After a top-paid arrival, their disruption falls by 0.022 and 0.020 in the first two periods (p = 0.29 and 0.41) and by 0.073 four to five years later (p = 0.006); pooled, it falls by 0.047 (p = 0.029). In both designs the pre-event period is close to the reference (0.013, p = 0.45, for retirements; 0.023, p = 0.30, for arrivals). Incumbents' citations rise after an arrival (pooled 0.067, p = 0.003; 0.060 and 0.062 in the two later periods, p = 0.015 and 0.012) but do not move after a retirement (0.002, p = 0.94). Productivity does not move in either design (−0.002, p = 0.84; 0.000, p = 0.98), nor do novelty (0.003, p = 0.84; 0.016, p = 0.58) or conventionality (−0.005, p = 0.87; 0.009, p = 0.78), although conventionality shows a marginal pre-event difference before retirements (p = 0.07). The two tests therefore find a within-department counterpart for disruption that is symmetric in direction, and one for citations after arrivals; because the department's measured dispersion changes only modestly, they cannot attribute these changes to dispersion alone. Although this design cannot establish a strict causal relationship between distributional regime and knowledge-production style, it provides useful supplementary evidence in data where departmental distribution regimes and cultures show limited year-to-year variation.

### 4.7. Alternative explanations

Our models include a stringent control set and, in the university-by-field specifications, compare departments within the same field and university. Still, the relationship among reward structure, collaborative structure, and research style may be co-evolutionary. We therefore discuss and examine five alternative explanations: team composition, selection at appointment, resource endowment, the consolidation of early breakthroughs, and the mix of sub-specializations inside departments.

**Team composition.** Unequal departments may also be departments of large teams, and large-team papers are mechanically lower on disruption and higher on citations (Wu et al. 2019). Table 2 could therefore reflect an aggregation artifact: unequal departments may retain scholars who lead large collaborations, whose papers have exactly the profile we observe, without pay dispersion bearing on anyone else.

We address this possibility in two steps. First, we rank scholars within each department by mean team size and drop the top 10 percent. The results survive: every coefficient keeps its sign and remains significant at the 5 percent level (disruption −0.029, $p = 0.039$). The citation and productivity estimates grow as the exclusion threshold rises, while the disruption, novelty and conventionality estimates stay close to their full-sample values. Second, we retain only the smallest-team half of each department: scholars who, by construction, do not lead its large collaborations. In this restricted sample, mean team size is 3.82 authors. The results are not weaker in this group (Table B1, Appendix B). In standardized terms the citation and productivity estimates are larger than in the full sample, at 0.053 ($p < 0.001$) and 0.063 ($p < 0.001$) against 0.049 and 0.040 in Table 2. Conventionality remains significant at 0.037 ($p = 0.038$); disruption (−0.026, $p = 0.054$) keeps its magnitude but is only marginally significant, and novelty (−0.023, $p = 0.13$) keeps its magnitude but not its significance.

The volume and impact results therefore also appear among scholars least involved in large collaborations. Although more unequal departments also tend to have larger teams, the pattern is more consistent with a department-level climate than with the aggregation of a few large-team scholars' outputs. The association also holds among small-team scholars.

**Selection at appointment.** Scientists who already do normal science may prefer, or be preferred by, departments with sharp internal pay competition. Nothing in Table 2 bears on this, since everything there is measured after arrival, so we move the question to the one juncture at which the two accounts separate: appointment. We identify scholars whose first appearance at a university is at assistant-professor rank and postdates that university's first year on record. Of 50,232 first appearances after a university's first observed year, 26,964 are at assistant rank, 21,591 sit in departments with their own pay equation and a departmental Gini for the year before the appointment, and 12,380 have at least three publications both before and after the appointment; each outcome is estimated on the appointees for whom it is observed both before and after (between 10,184 and 12,373, depending on the outcome), so the pre- and post-hire estimates describe the same people. For each such individual, we compute the profile the person already had, aggregating only over works published strictly before arrival, and the profile after arrival from the works published afterwards, and regress each on the department's residual Gini in the preceding year, with Table 2's fixed effects, transform, and clustering and with controls for career age at appointment and its square, gender, race, the department's female and minority shares, log department size, and the appointment year; results appear in Appendix B Table B2. [4]

The pre-hire record differs across the five outcomes. Departments with greater unexplained pay dispersion show no detectable tendency to appoint junior scientists whose prior work is more disruptive (in standardized terms 0.007, $p = 0.66$; 0.015, $p = 0.44$ under university-by-field fixed effects), more cited (0.007, $p = 0.64$; 0.007, $p = 0.67$), or more productive (0.007, $p = 0.59$; 0.005, $p = 0.76$). Their appointees' prior work is at most marginally more conventional (0.023, $p = 0.100$; 0.022, $p = 0.18$) and not detectably less novel (−0.014, $p = 0.23$; −0.017, $p = 0.23$). After arrival, the work of these same scholars is more conventional (0.037, $p = 0.016$; 0.034, $p = 0.047$) and less novel (−0.040, $p = 0.010$; −0.035, $p = 0.014$). Selection at appointment may therefore account for part of the novelty and conventionality results, although the pre-hire evidence for it is weak. We therefore find no evidence that the disruption, novelty, citation, and productivity results are recruitment patterns, whereas for conventionality the record scholars bring with them shows weak signs of selection.

**Early breakthrough, later consolidation.** A third explanation is that high salaries reward early breakthroughs, after which scholars shift toward less disruptive work in larger teams. The pattern in Table 2 would therefore reflect the later careers of scientists rewarded for earlier achievements. Two findings challenge this account (Appendix Table B4).

First, a more disruptive record is associated with lower annual pay growth at every career stage. Re-estimating the model in Appendix Table B3 separately for early-, mid-, and late-career scholars yields coefficients of −0.50, −0.41, and −0.34 percentage points per rank-normal unit, respectively. Using only the preceding three years of publications produces smaller but still negative estimates (−0.03, −0.11, and −0.15), whereas productivity is rewarded by pay growth throughout. Therefore, disruptive works are not rewarded through annual pay growth in early career.

Second, among scholars observed at career ages ≤10 and >15, early disruption predicts a lower later pay premium, holding other dimensions of the initial publication record fixed ($\beta = -0.0126$, $p = 0.01$). This association persists after controlling for initial compensation above the departmental ladder. Early disruption also predicts smaller later teams ($\beta = -0.022$) and more disruptive subsequent work ($\beta = 0.015$), while early productivity predicts a higher future pay premium ($\beta = 0.0238$). These results provide little support for a trajectory in which rewarded early disruptors later turn toward less disruptive research in larger collaborations; rather, they show that people's research styles tend to be constant across time.

**Sub-specialization mix.** Departments differ in the concentration of intellectual specializations, and scholars from different subfields differ in how they publish and cite: comparative-historical sociologists may cite more and publish less than stratification scholars. A department's mix could then produce the Table 2 pattern. To examine this explanation, we assign each scholar the most frequent OpenAlex subfield/topic among the works they published while in the department (in our data, we identified 239 of the taxonomy's 252 subfields and 3,631 of its 4,516 topics), and summarize each department's domain distribution by four measures: the number of distinct categories, the Herfindahl of their shares, Shannon entropy and the share in the most common category (Appendix Table B5). As expected, more unequal departments do house a more diverse mix (a one-SD larger log number of subfields goes with a 0.107-SD higher residualized Gini, and a one-SD larger log number of topics with a 0.158-SD higher one, department size controlled), but the mix does not account for the results.

Adding the four composition measures to Table 2 preserves most coefficients' signs and significance, but weakens productivity to marginal significance at the subfield level (0.033, $p < 0.10$) and below conventional thresholds at the topic level (0.023, not significant) (Table B5, Appendix B, Panel A/B). Comparing scholars within their own subfield or topic likewise leaves novelty, conventionality, and disruption significant in the same directions. The citation estimate remains positive and significant but becomes smaller, whereas productivity is no longer statistically distinguishable from zero. Thus, the findings on the direction of work (H2) are robust to the mix of sub-specializations, while those on volume (H1), especially productivity, are partly compositional.

**Resource endowment.** Departments with well-funded, heavily cited faculty may have both the conditions that generate unexplained pay dispersion, such as outside offers, retention counters, and soft-money arrangements, and the research programs that produce high-volume, highly cited, large-team science. In that case, unexplained dispersion and normal science would be joint products of a department's position in a steeply sorted labor market, where the arrival or departure of a single star can reshape the output of everyone nearby (Azoulay, Graff Zivin, and Wang 2010; Clauset, Arbesman, and Larremore 2015; Way et al. 2019; Wapman et al. 2022).

We address this possibility partly through model specification. Every model controls for log department size, female share, race minority share, and absorbs university and field fixed effects jointly. The matched design also balances on size quintile, seniority band, broad area, and institutional ranking status. The first stage does not net out h-index or grant income, so a department that pays its most-funded people more is, by design, recorded as more unequal. But when we add those terms back into each department's pay equation, the estimates keep their signs and similar magnitudes, although the disruption estimate is only marginally significant ($p = 0.053$; Section 4.5). The association therefore does not operate through pay for measured performance alone.

What we cannot directly control is the department's endowment or aggregate grant income. We therefore regard this explanation as reduced but not eliminated. We discuss it as a limitation.

## 5. Discussion

Tournament theory holds that a wider spread of rewards elicits greater effort (Lazear and Rosen 1981), while comparison and organizational justice traditions argue that unequal rewards invite judgments of legitimacy that can weaken motivation (Adams 1965; Pfeffer and Langton 1993). We have argued that both accounts are incomplete for knowledge work, because some of its value is not legible under prevailing standards.

We examine this argument in a setting where both components of output can be measured at population scale. Using a department-specific pay equation, we isolate the pay dispersion that each department's formal ladder cannot explain. Publication rates and citation counts are higher in departments with more of that dispersion, as tournament theory would predict. But work in those same departments is less disruptive, less novel, and more conventional. Three interpretive points follow from these patterns.

**First, the Kuhnian distinction clarifies what standard accounts of pay dispersion miss.** In knowledge work, evaluation depends on signals that are legible within the current intellectual frame. Organizations can reward what they can see, and what they can see is usually work that extends existing standards rather than work that changes them.

The implication is sociological as much as economic. Rewards do not simply motivate individuals; they help define which forms of contribution become recognizable as valuable. When evaluation is organized around existing standards, it strengthens the social order of the field by directing attention and resources toward work that already fits that order. The individual premium results make this concrete. What departments pay above their formal ladder is a premium for scale and breadth, the visible profile of the productive scholar. It does not reward departure from the frame. As a result, ideas and knowledge with the potential to disrupt existing systems, socially or economically, may become less likely where inequality is greater.

Interestingly, we also observe a positive association between disruption and promotion from assistant to associate professor (Appendix Table B3). One possible interpretation is that departments recognize the conservative tilt of their pay systems and use promotion as another way to encourage innovation: advancement may favor disruptive work even as pay premiums reward scale and breadth. This pattern does not undermine our broader argument that reward systems can shape knowledge production by influencing what scholars study and how they work on them.

Second, the correspondence appears to be a property of departments rather than of the scholars they initially hire, which is where relational accounts of organizational inequality would locate it. More unequal departments show no detectable tendency to recruit junior scholars whose prior work is more disruptive, highly cited, or productive, although weak pre-hire differences in conventionality suggest some selection. Among incumbents, disruptive work rises modestly after a favored senior member leaves and declines after a top-paid colleague arrives. While our data cannot causally determine whether this pattern reflects the effect of reward structures on incumbents or the continued sorting of scholars into and out of departments, it points to the department as the unit that prices people relationally (Avent-Holt and Tomaskovic-Devey 2019; Tomaskovic-Devey and Avent-Holt 2019). It also connects to the sorting of scientists across departments that the science of science has documented from the other side (Clauset et al. 2015; Way et al. 2019; Wapman et al. 2022).

Departments differ not only in prestige and resources, but also in how much of their pay dispersion can be explained by formal ladders and merit-based metrics, and in which merit metrics they choose to recognize. These differences align with the kind of science their members produce. The sociological reading of value, as a set of claims about worth that must be evaluated as legitimate, fits this evidence better than the view of dispersion as a simple prize schedule. We suggest, however, that this process operates at the collective level as well as the individual level. The relevant comparison is not only a scholar deciding how hard to work, but a department deciding how to reward its members.

**Third, the operative unit is the department, and its dispersion is not designed.** Although the university-by-field specifications compare departments in the same field at the same university, the identifying variation still comes from organizational units that may differ in unobserved ways. Unexplained dispersion accumulates through retention counters, outside-offer matches, endowed differentials, and soft-money arrangements. Each may be locally defensible, but the distribution they jointly produce is rarely subject to collective decision and may reflect both noise and selection. Our results therefore should not be interpreted as strictly causal. They instead reveal a meaningful association that should matter to scholars and policymakers alike.

**Limitations.** Like all studies, ours has limitations. First, the estimates remain comparisons between departments, and a department's unexplained pay dispersion is not assigned. The appointment test narrows the interpretation only in part: it detects no statistically significant differences in the pre-hire record for disruption, novelty, citations, or productivity, so the pattern is not visibly a recruitment pattern. The first-stage checks also rule out the most direct construction artifacts. Still, none of these tests identifies a strict causal effect. The resource-endowment account, which we cannot fully rule out, could generate the volume and citation associations without any behavioral response. Resolving this would require variation in departmental pay regimes generated outside the department, such as a state compensation reform or a change in collective bargaining, which goes beyond our data here.

Second, our coverage is limited. The dataset includes only U.S. public universities with disclosure statutes, and only for the years they disclose. Private institutions are absent, and the sample is weighted toward large public research universities. We think the data can nevertheless support the research question here, as the mechanism is defined at the department level, and in these records departments enter or leave as units rather than as selected individuals. What the design gives up, however, is range. Private universities likely sustain larger unexplained pay dispersion, and public-sector pay is more compressed at the top. The upper tail of the reward structure most relevant to the theory is therefore thinner here than in the full population of research universities. The estimates may not generalize to the wider range of reward structures at private universities; the direction of any resulting difference cannot be established from these data.

**6. Conclusion**

We began with a question that the sociology of science usually asks in one direction: how does knowledge production shape the stratification of scientists? We ask the reverse question: how does stratification shape the kind of knowledge scientists produce?

Departments with more pay dispersion unexplained by their own formal ladders have scholars who publish more, receive more citations, use prior literature more conventionally, and produce work that is less likely to displace what came before. Citation and productivity coefficients retain their positive signs across the reported alternatives, but productivity is not statistically distinguishable from zero in the CEM sample and in several alternative specifications. The disruption result is moderately robust. Novelty and conventionality hold between departments, survive matching and various controls, and are associated with both residualized and raw pay inequality. While our evidence does not demonstrate that pay dispersion causes the pattern, we do find a consistent and stable association between reward structure and knowledge production across departments, with a within-department counterpart for disruption when a top-paid senior member leaves or arrives.

The broader theoretical claim is that inequality can carry an innovation cost. Incentive theories, including tournament theory, do not translate straightforwardly to science and other forms of creative work. The issue is not that knowledge workers are indifferent to pay, or that pay dispersion simply demotivates them; the productivity and citation results are inconsistent with those interpretations. The issue is that rewards depend on signals available at the moment of evaluation. Existing metrics are better suited to recognizing work that extends the current order than work that revises it. In creative work, those signals are always incomplete because the work depends on standards and values that are not yet socially settled. Reward systems may therefore amplify what is already legible while discounting disruptive contributions.

We test the argument in academic science because both modes of knowledge production can be measured at population scale there, not because the argument is specific to science. Under the scope conditions described above, the same compositional prediction should apply to other creative professions, including corporate research and development, design, and other forms of knowledge-intensive work. It may also extend to socioeconomic settings in which value and order are not yet settled, and where decisions are made under that ambiguity (Stark 2009). In such settings, our argument predicts that growing inequality may favor more conservative choices and reduce the likelihood of work that disrupts the existing system.

This paper also makes two methodological contributions. First, by estimating a separate pay equation for each organizational unit, we construct two theoretically distinct measures: ladder-based or merit-based pay inequality, and residual dispersion unexplained by that ladder. Although we do not pursue this distinction further here, it offers a useful direction for future research in economics and organizational sociology. Second, we build an AI-agent pipeline that links two decades of public-university payroll records to OpenAlex at the personal level. To our knowledge, this is currently one of the largest panels of its kind.

The practical implication follows from the volume and impact results alone. Universities that widen internal pay dispersion in pursuit of research excellence may see gains in the outcomes directly observable: more publications and higher citations. Yet these departments are less likely to be home to truly disruptive work, which challenges established knowledge and reshapes how contributions are valued. Our design cannot establish whether widening pay gaps would itself cause this pattern, but we do observe consistent associations. Whether this is a favorable tradeoff is not a question our data can answer. It is a tradeoff nonetheless, and measuring only one side of it is not neutral.

**Notes**

1. The department-specific first stage can be estimated only for departments with enough residual degrees of freedom. We set this threshold at five residual degrees of freedom after estimating an intercept, up to six slope parameters, and T−1 year effects, with constant or collinear columns omitted; 4,839 of 5,549 departments meet that criterion. Departments below the threshold are smaller on average. Figure 3 shows that including them with pooled residuals attenuates every coefficient toward zero but leaves all signs unchanged.

2. The leverage correction $\tilde{e}_i = \frac{e_i}{\sqrt{1 - h_{ii}}}$ follows the standardized-residual adjustment familiar from regression diagnostics, but its purpose here is definitional rather than diagnostic. Without it, the residual Gini mechanically declines with department size, even as department size enters every second-stage model as a control.

3. In both the retirement and arrival designs, we rank faculty within each department-year by residual pay premium, from highest to lowest. Ties receive the best rank they occupy. The top-premium group includes those with a positive premium and rank $r \leq \lceil 0.1n \rceil$, where $n$ is department-year headcount. This rule includes ties at the cutoff; when $n \leq 10$, only the highest-premium member or tied members qualify. We cap outcome years at the point where the underlying indices become unreliable: disruption, novelty, conventionality, and citations are observed through 2023, and productivity through 2025. Because departures to private, foreign, or non-academic employers cannot be distinguished from retirement, the career-age restriction carries most of the design. The arrival design dates a scholar's arrival from their first appearance in a university's payroll records after the university's first observed year, as in the appointment test.

4. The appointment test dates a scholar's arrival from their first appearance in a university's payroll records and requires that year to fall after the university's first observed year. This rule misses scholars hired in the first year a university enters the data and may misclassify returning faculty as new hires; both can affect sample selection and appointment timing, with an uncertain direction of bias.

## Data Availability Statement

The salary records used in this article come from public disclosures released by U.S. state agencies and public universities under state transparency statutes and may be redistributed. Publication and citation information comes from OpenAlex, which is openly licensed.

**Reproducibility package.** All data, code, and documentation needed to reproduce the empirical results in this paper are available at https://doi.org/10.5281/zenodo.23645690. Names of universities, departments, and individuals are replaced by codes, and the author retains the mapping to the original identifiers. The estimation frames and analysis code included in the package reproduce every table, figure, and reported number. A step-by-step replication guide in Markdown is also provided to support automated, machine-assisted replication.

**AI Disclosure.** An AI agent, Claude Code (Fable 5.1, Anthropic), wrote and ran scripts to collect public salary data from multiple online sources, as described in the method section. It also helped audit, clean, and assemble the data and assess ambiguous cases. Codex (GPT6-Astra) assisted with code review, auditing, and language editing. Auditing scripts and agents had no access to the collection cache.

The author designed the study, developed the prompts, specified the statistical models, interpreted the results, and wrote the manuscript. I reviewed the Python code and final text, verified the content, and took full responsibility for the article.

## References

Adams, J. Stacy. 1965. "Inequity in Social Exchange." Pp. 267–299 in *Advances in Experimental Social Psychology*. Vol. 2, edited by Leonard Berkowitz. New York: Academic Press.

Akerlof, George A., and Janet L. Yellen. 1990. "The Fair Wage-Effort Hypothesis and Unemployment." *Quarterly Journal of Economics* 105(2):255–283.

Amabile, Teresa M. 1996. *Creativity in Context*. Boulder, CO: Westview Press.

Avent-Holt, Dustin, and Donald Tomaskovic-Devey. 2019. "Organizations as the Building Blocks of Social Inequalities." *Sociology Compass* 13(2):e12655.

Azoulay, Pierre, Joshua S. Graff Zivin, and Gustavo Manso. 2011. "Incentives and Creativity: Evidence from the Academic Life Sciences." *RAND Journal of Economics* 42(3):527–554.

Azoulay, Pierre, Joshua S. Graff Zivin, and Jialan Wang. 2010. "Superstar Extinction." *Quarterly Journal of Economics* 125(2):549–589.

Baker, George P., Michael C. Jensen, and Kevin J. Murphy. 1988. "Compensation and Incentives: Practice vs. Theory." *Journal of Finance* 43(3):593–616.

Bol, Thijs, Mathijs de Vaan, and Arnout van de Rijt. 2018. "The Matthew Effect in Science Funding." *Proceedings of the National Academy of Sciences* 115(19):4887–4890.

Bornmann, Lutz, and Alexander Tekles. 2019. "Disruption Index Depends on Length of Citation Window." *El Profesional de la Información* 28(2):e280207.

Bornmann, Lutz, Alexander Tekles, Helena H. Zhang, and Fred Y. Ye. 2019. "Do We Measure Novelty When We Analyze Unusual Combinations of Cited References? A Validation Study of Bibliometric Novelty Indicators Based on F1000Prime Data." *Journal of Informetrics* 13(4):100979.

Boudreau, Kevin J., Eva C. Guinan, Karim R. Lakhani, and Christoph Riedl. 2016. "Looking Across and Looking Beyond the Knowledge Frontier: Intellectual Distance, Novelty, and Resource Allocation in Science." *Management Science* 62(10):2765–2783.

Breza, Emily, Supreet Kaur, and Yogita Shamdasani. 2018. "The Morale Effects of Pay Inequality." *Quarterly Journal of Economics* 133(2):611–663.

Bull, Clive, Andrew Schotter, and Keith Weigelt. 1987. "Tournaments and Piece Rates: An Experimental Study." *Journal of Political Economy* 95(1):1–33.

Card, David, Alexandre Mas, Enrico Moretti, and Emmanuel Saez. 2012. "Inequality at Work: The Effect of Peer Salaries on Job Satisfaction." *American Economic Review* 102(6):2981–3003.

Castilla, Emilio J. 2008. "Gender, Race, and Meritocracy in Organizational Careers." *American Journal of Sociology* 113(6):1479–1526.

Castilla, Emilio J. 2015. "Accounting for the Gap: A Firm Study Manipulating Organizational Accountability and Transparency in Pay Decisions." *Organization Science* 26(2):311–333.

Chen, Kong-Pin. 2003. "Sabotage in Promotion Tournaments." *Journal of Law, Economics, and Organization* 19(1):119–140.

Clauset, Aaron, Samuel Arbesman, and Daniel B. Larremore. 2015. "Systematic Inequality and Hierarchy in Faculty Hiring Networks." *Science Advances* 1(1):e1400005.

Cobb, J. Adam, and Flannery G. Stevens. 2017. "These Unequal States: Corporate Organization and Income Inequality in the United States." *Administrative Science Quarterly* 62(2):304–340.

Cole, Jonathan R., and Stephen Cole. 1973. *Social Stratification in Science*. Chicago: University of Chicago Press.

Colquitt, Jason A., Donald E. Conlon, Michael J. Wesson, Christopher O. L. H. Porter, and K. Yee Ng. 2001. "Justice at the Millennium: A Meta-Analytic Review of 25 Years of Organizational Justice Research." *Journal of Applied Psychology* 86(3):425–445.

Connelly, Brian L., Laszlo Tihanyi, T. Russell Crook, and K. Ashley Gangloff. 2014. "Tournament Theory: Thirty Years of Contests and Competitions." *Journal of Management* 40(1):16–47.

Dahler-Larsen, Peter. 2014. "Constitutive Effects of Performance Indicators: Getting beyond Unintended Consequences." *Public Management Review* 16(7):969–986.

Dannefer, Dale. 2003. "Cumulative Advantage/Disadvantage and the Life Course: Cross-Fertilizing Age and Social Science Theory." *Journals of Gerontology, Series B: Psychological Sciences and Social Sciences* 58(6):S327–S337.

de Rijcke, Sarah, Paul F. Wouters, Alex D. Rushforth, Thomas P. Franssen, and Björn Hammarfelt. 2016. "Evaluation Practices and Effects of Indicator Use—A Literature Review." *Research Evaluation* 25(2):161–169.

Deutsch, Morton. 1985. *Distributive Justice: A Social-Psychological Perspective*. New Haven, CT: Yale University Press.

Ederer, Florian, and Gustavo Manso. 2013. "Is Pay for Performance Detrimental to Innovation?" *Management Science* 59(7):1496–1513.

Ehrenberg, Ronald G., and Michael L. Bognanno. 1990. "Do Tournaments Have Incentive Effects?" *Journal of Political Economy* 98(6):1307–1324.

Espeland, Wendy Nelson, and Michael Sauder. 2007. "Rankings and Reactivity: How Public Measures Recreate Social Worlds." *American Journal of Sociology* 113(1):1–40.

Espeland, Wendy Nelson, and Mitchell L. Stevens. 1998. "Commensuration as a Social Process." *Annual Review of Sociology* 24:313–343.

Fehr, Ernst, and Klaus M. Schmidt. 1999. "A Theory of Fairness, Competition, and Cooperation." *Quarterly Journal of Economics* 114(3):817–868.

Festinger, Leon. 1954. "A Theory of Social Comparison Processes." *Human Relations* 7(2):117–140.

Fortunato, Santo, Carl T. Bergstrom, Katy Börner, James A. Evans, Dirk Helbing, Staša Milojević, Alexander M. Petersen, Filippo Radicchi, Roberta Sinatra, Brian Uzzi, Alessandro Vespignani, Ludo Waltman, Dashun Wang, and Albert-László Barabási. 2018. "Science of Science." *Science* 359(6379):eaao0185.

Foster, Jacob G., Andrey Rzhetsky, and James A. Evans. 2015. "Tradition and Innovation in Scientists' Research Strategies." *American Sociological Review* 80(5):875–908.

Frey, Bruno S., and Reto Jegen. 2001. "Motivation Crowding Theory." *Journal of Economic Surveys* 15(5):589–611.

Funk, Russell J., and Jason Owen-Smith. 2017. "A Dynamic Network Measure of Technological Change." *Management Science* 63(3):791–817.

Holmström, Bengt, and Paul Milgrom. 1991. "Multitask Principal–Agent Analyses: Incentive Contracts, Asset Ownership, and Job Design." *Journal of Law, Economics, and Organization* 7(special issue):24–52.

Holst, Vincent, Andres Algaba, Floriano Tori, Sylvia Wenmackers, and Vincent Ginis. 2026. "Dataset Artefacts Can Partially Drive the Measured Decline in Disruption." *Nature* 656(8127):E7–E13.

Iacus, Stefano M., Gary King, and Giuseppe Porro. 2011. "Multivariate Matching Methods That Are Monotonic Imbalance Bounding." *Journal of the American Statistical Association* 106(493):345–361.

Iacus, Stefano M., Gary King, and Giuseppe Porro. 2012. "Causal Inference without Balance Checking: Coarsened Exact Matching." *Political Analysis* 20(1):1–24.

Kacperczyk, Aleksandra, and Chanchal Balachandran. 2018. "Vertical and Horizontal Wage Dispersion and Mobility Outcomes: Evidence from the Swedish Microdata." *Organization Science* 29(1):17–38.

Ke, Qing, Emilio Ferrara, Filippo Radicchi, and Alessandro Flammini. 2015. "Defining and Identifying Sleeping Beauties in Science." *Proceedings of the National Academy of Sciences* 112(24):7426–7431.

Kuhn, Thomas S. 1962. *The Structure of Scientific Revolutions*. Chicago: University of Chicago Press.

Kuhn, Thomas S. 1977. *The Essential Tension: Selected Studies in Scientific Tradition and Change*. Chicago: University of Chicago Press.

Larivière, Vincent, Yves Gingras, Cassidy R. Sugimoto, and Andrew Tsou. 2015. "Team Size Matters: Collaboration and Scientific Impact since 1900." *Journal of the Association for Information Science and Technology* 66(7):1323–1332.

Lazear, Edward P., and Sherwin Rosen. 1981. "Rank-Order Tournaments as Optimum Labor Contracts." *Journal of Political Economy* 89(5):841–864.

Leahey, Erin, Jina Lee, and Russell J. Funk. 2023. "What Types of Novelty Are Most Disruptive?" *American Sociological Review* 88(3):562–597.

Leibel, Christian, and Lutz Bornmann. 2024. "What Do We Know about the Disruption Index in Scientometrics? An Overview of the Literature." *Scientometrics* 129(1):601–639.

Lerner, Josh, and Julie Wulf. 2007. "Innovation and Incentives: Evidence from Corporate R&D." *Review of Economics and Statistics* 89(4):634–644.

Lind, E. Allan, and Tom R. Tyler. 1988. *The Social Psychology of Procedural Justice*. New York: Plenum Press.

Lockhart, Jeffrey W., Molly M. King, and Christin Munsch. 2023. "Name-Based Demographic Inference and the Unequal Distribution of Misrecognition." *Nature Human Behaviour* 7(7):1084–1095.

Manso, Gustavo. 2011. "Motivating Innovation." *Journal of Finance* 66(5):1823–1860.

Merton, Robert K. 1968. "The Matthew Effect in Science." *Science* 159(3810):56–63.

Merton, Robert K. 1988. "The Matthew Effect in Science, II: Cumulative Advantage and the Symbolism of Intellectual Property." *Isis* 79(4):606–623.

O'Reilly, Charles A., III, Brian G. Main, and Graef S. Crystal. 1988. "CEO Compensation as Tournament and Social Comparison: A Tale of Two Theories." *Administrative Science Quarterly* 33(2):257–274.

Park, Michael, Erin Leahey, and Russell J. Funk. 2023. "Papers and Patents Are Becoming Less Disruptive over Time." *Nature* 613(7942):138–144.

Park, Michael, Erin Leahey, and Russell J. Funk. 2026. "Reply to: Dataset Artefacts Can Partially Drive the Measured Decline in Disruption." *Nature* 656(8127):E14–E21.

Petersen, Alexander M., Felber Arroyave, and Fabio Pammolli. 2024. "The Disruption Index Is Biased by Citation Inflation." *Quantitative Science Studies* 5(4):936–953.

Pfeffer, Jeffrey, and Nancy Langton. 1993. “The Effect of Wage Dispersion on Satisfaction, Productivity, and Working Collaboratively: Evidence from College and University Faculty.” *Administrative Science Quarterly* 38(3):382–407.
Prendergast, Canice. 1999. “The Provision of Incentives in Firms.” *Journal of Economic Literature* 37(1):7–63.
Priem, Jason, Heather Piwowar, and Richard Orr. 2022. “OpenAlex: A Fully-Open Index of Scholarly Works, Authors, Venues, Institutions, and Concepts.” arXiv:2205.01833.
Sauder, Michael, and Wendy Nelson Espeland. 2009. “The Discipline of Rankings: Tight Coupling and Organizational Change.” *American Sociological Review* 74(1):63–82.
Shaw, Jason D. 2014. “Pay Dispersion.” *Annual Review of Organizational Psychology and Organizational Behavior* 1:521–544.
Shaw, Jason D., Nina Gupta, and John E. Delery. 2002. “Pay Dispersion and Workforce Performance: Moderating Effects of Incentives and Interdependence.” *Strategic Management Journal* 23(6):491–512.
Sinatra, Roberta, Dashun Wang, Pierre Deville, Chaoming Song, and Albert-László Barabási. 2016. “Quantifying the Evolution of Individual Scientific Impact.” *Science* 354(6312):aaf5239.
Stark, David. 2009. *The Sense of Dissonance: Accounts of Worth in Economic Life*. Princeton, NJ: Princeton University Press.
Tomaskovic-Devey, Donald, and Dustin Avent-Holt. 2019. *Relational Inequalities: An Organizational Approach*. New York: Oxford University Press.
Uzzi, Brian, Satyam Mukherjee, Michael Stringer, and Ben Jones. 2013. “Atypical Combinations and Scientific Impact.” *Science* 342(6157):468–472.
Wang, Dashun, and Albert-László Barabási. 2021. *The Science of Science*. Cambridge, UK: Cambridge University Press.
Wang, Jian, Reinhilde Veugelers, and Paula Stephan. 2017. “Bias against Novelty in Science: A Cautionary Tale for Users of Bibliometric Indicators.” *Research Policy* 46(8):1416–1436.
Wang, Kuansan, Zhihong Shen, Chiyuan Huang, Chieh-Han Wu, Yuxiao Dong, and Anshul Kanakia. 2020. “Microsoft Academic Graph: When Experts Are Not Enough.” *Quantitative Science Studies* 1(1):396–413.
Wapman, K. Hunter, Sam Zhang, Aaron Clauset, and Daniel B. Larremore. 2022. “Quantifying Hierarchy and Dynamics in US Faculty Hiring and Retention.” *Nature* 610(7930):120–127.
Way, Samuel F., Allison C. Morgan, Daniel B. Larremore, and Aaron Clauset. 2019. “Productivity, Prominence, and the Effects of Academic Environment.” *Proceedings of the National Academy of Sciences* 116(22):10729–10733.
Wu, Lingfei, Dashun Wang, and James A. Evans. 2019. “Large Teams Develop and Small Teams Disrupt Science and Technology.” *Nature* 566(7744):378–382.
Wuchty, Stefan, Benjamin F. Jones, and Brian Uzzi. 2007. “The Increasing Dominance of Teams in Production of Knowledge.” *Science* 316(5827):1036–1039.
Zuckerman, Harriet. 1977. *Scientific Elite: Nobel Laureates in the United States*. New York: Free Press.